\documentclass{article}
\usepackage{spconf,amsmath,graphicx}
\usepackage{graphicx}
\usepackage{subcaption}
\usepackage{booktabs}
\usepackage{siunitx}
\usepackage{amssymb}
\usepackage{rotating}
\usepackage{makecell}
\usepackage{hyperref}

\title{DualSep: A Light-weight Dual-Encoder Convolutional Recurrent Network For Real-Time In-Car Speech Separation}
\name{Ziqian Wang$^1$, Jiayao Sun$^1$, Zihan Zhang$^1$, Xingchen Li$^1$, Jie Liu$^2$, Lei Xie$^1$$^{*}$\thanks{* Corresponding author.}}
\address{
$^1$Audio, Speech and Language Processing Group (ASLP@NPU), School of Computer Science, \\Northwestern Polytechnical University, Xi'an, China\\
$^2$Huawei Cloud}

\begin{document}
%
\maketitle
\begin{abstract}
Advancements in deep learning and voice-activated technologies have driven the development of human-vehicle interaction. Distributed microphone arrays are widely used in in-car scenarios because they can accurately capture the voices of passengers from different speech zones. However, the increase in the number of audio channels, coupled with the limited computational resources and low latency requirements of in-car systems, presents challenges for in-car multi-channel speech separation. To migrate the problems, we propose a lightweight framework that cascades digital signal processing (DSP) and neural networks (NN). We utilize fixed beamforming (BF) to reduce computational costs and independent vector analysis (IVA) to provide spatial prior. We employ dual encoders for dual-branch modeling, with spatial encoder capturing spatial cues and spectral encoder preserving spectral information, facilitating spatial-spectral fusion. Our proposed system supports both streaming and non-streaming modes. Experimental results demonstrate the superiority of the proposed system across various metrics. With only 0.83M parameters and 0.39 real-time factor (RTF) on an Intel Core i7 (2.6GHz) CPU, it effectively separates speech into distinct speech zones. Our demos are available at \href{https://honee-w.github.io/DualSep/}{https://honee-w.github.io/DualSep/}.
\end{abstract}
\section{Introduction}
\label{sec:intro}


In recent years, with the development of deep learning, technologies such as automatic speech recognition (ASR)~\cite{zhang2022end}, speech enhancement (SE)~\cite{wang2020complex}, and speaker identification (SI)~\cite{kanda2020joint} have rapidly advanced, significantly improving the human-machine interaction experience~\cite{pickering2007research, rahmati2020game,capallera2022human, murali2022intelligent}. As smart devices evolve, there is a growing demand for more natural and convenient interactions, such as human-vehicle voice interaction. When there are multiple passengers in the car, speech separation is a crucial technology to ensure accurate interaction. However, the in-car environment poses unique challenges for effective speech separation: the use of distributed microphone arrays to better capture the voices of passengers in different zones increases the number of audio channels, the limited computational resources and low-latency requirements within the car, and the various noises that may occur during driving. Overcoming these challenges is essential for downstream applications that rely on accurate in-car speech separation, such as ASR and spoken language processing.

Currently, deep learning-based speech separation systems have demonstrated remarkable performances, leading to enhanced speech quality and intelligibility. 
In multichannel scenarios, these systems primarily fall into two categories.
The first category utilizes neural networks as neural beamformers, directly inputting multichannel signals to learn weights for filter-and-sum operations implicitly in an end-to-end fashion~\cite{xu2021generalized, wang2021multi, zhang2022all}. This approach has proven effective in achieving superior speech quality by reducing noise or interfering speech. However, it can encounter degradation in real-world scenarios due to overfitting to testsets.
Concurrently, the second category adopts a multichannel processing scheme, employing neural networks to explicitly derive beamforming filtering weights from estimated speech and noise components~\cite{zhang2021adl}. This includes methods such as the multichannel wiener filter (MWF)~\cite{van2009speech}, 
the generalized eigenvalue beamformer (GEV)~\cite{xu2021generalized}, and the minimum variance distortionless response (MVDR) filter~\cite{zhang2021adl, kothapally2023deep}, among others. These approaches can alleviate nonlinear distortions and provide advantages for downstream ASR systems. Nonetheless, the involved matrix operations (e.g., matrix inversion) are sometimes numerically unstable when jointly trained with neural networks~\cite{zhao2012fast} and they can suffer from residual noise issues~\cite{ xiao2017time, boeddeker2018exploring, xu2020neural}.

Besides their effectiveness, the above systems share a common issue, they often utilize non-causal convolution or bidirectional recurrent neural networks that incorporate future or global information for contextual modeling to obtain better performances. However, such approaches are unsuitable for real-world streaming scenarios due to large model size, slow inference time, and high computational overhead.

To address these challenges, we propose a lightweight multi-channel speech separation system that combines DSP and NN, where DSP acts as a preprocessor and NN acts as a separator and refiner. Specifically, our proposed system initially utilizes fixed beamforming (BF)~\cite{gannot2017consolidated} to reduce computational overhead and IVA~\cite{kim2006independent} to provide spatial prior. Subsequently, we employ dual encoders~\cite{chidambaram2018learning} to enable dual-branch modeling. The spatial encoder is designed to capture spatial cues, while the spectral encoder focuses on preserving spectral information, facilitating spatial-spectral fusion. To better model multi-dimensional context, we then integrate the triple-path modeling module to capture the spectral, spatial, and temporal characteristics within latent representations. Finally, the decoder reconstructs the separated speech. 
The proposed system supports both streaming and non-streaming modes, with the streaming mode achieving performance close to that of non-streaming.
Experimental results show that our proposed system achieves the best performance, 
effectively separating speech into distinct speech zones. With only 0.83M parameters and 0.39 RTF on an Intel Core i7 (2.6GHz) CPU, it enhances speech quality and improves recognition rates for downstream ASR systems, holding promise for real-time communication and offline applications.

\section{Problem Formulation}
\label{sec:problem}

\begin{figure*}[]
  \centering
  \includegraphics[width=0.95\linewidth]{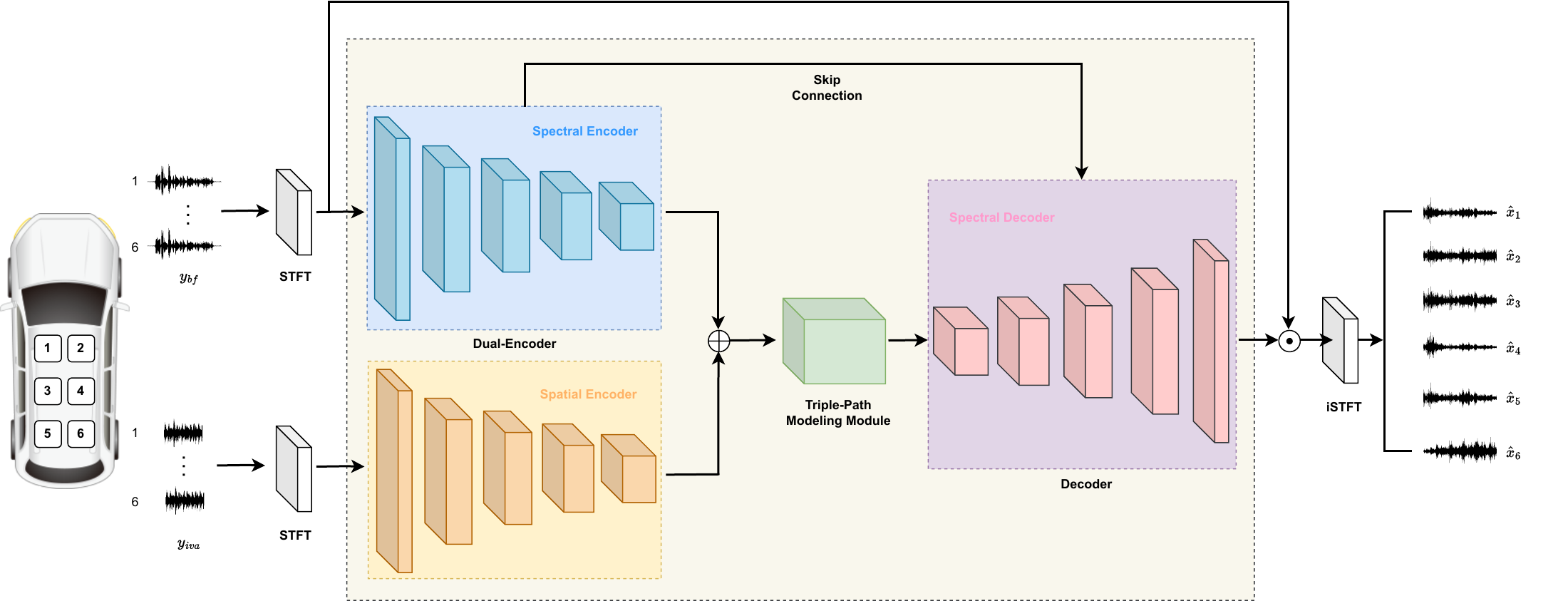}
\caption{Overview of the proposed DualSep  Featuring a dual-encoder, a triple-path modeling module, and a decoder, the system processes beamformed and IVA-enhanced inputs, modeling spatial priors and feature dependencies across time, frequency, and channels to produce separated speech for each in-car speech zone.}
\label{fig:pipeline}
\end{figure*}

We address the challenge of in-car speech separation with $N$ speakers by deploying a $P$-channel microphone array in each of the $M$ distinct speech zones ($N \le M$). Let $s_{i}(n)$ denote the clean speech from the $i$-th speaker. Assuming a designated zone where only one speaker can move while speaking (see Fig. \ref{fig:pipeline} for speaker zones), the recorded speech by the microphone array, denoted as $y(n)$, can be expressed as:
\begin{align}
\label{eq:formulation}
x_{i}(n) &= r_{i}(n) * s_{i}(n), \\
y(n) &= \sum_{i=0}^{N-1} x_{i}(n) + v(n), 
\end{align}
where, $n$ is the sample index, $r_{i}(n)$ represents the $P\times M$-channel room impulse responses (RIRs) from the $i$-th speaker to the microphone arrays, $*$ denotes convolution operation, and $v(n)$ represents background noise. Our objective is to develop a lightweight speech separation system ($\Theta_{\text{proposed}}$) that can effectively separate all speakers' speech while simultaneously suppressing background noise with lower computational costs and faster inference, see Eq. (\ref{eq:target}). The vector $\hat{x}(n)$ comprises all  $\hat{x}_{i}(n), \forall i \in[0, N-1]$. It is essential to note that the proposed separation system focuses on estimating reverberant clean speech rather than the anechoic speech $\hat{s}(n)$.
\begin{equation}
\label{eq:target}
\hat{x}(n) = \Theta _{\text{proposed}}(y(n))
\end{equation}

\section{Proposed System}
\label{sec:propose}

As illustrated in Fig. \ref{fig:pipeline},  the proposed DualSep is built on an encoder-decoder framework, featuring a dual-encoder, a triple-path modeling module, and a decoder. 
Two input signals, $y_{bf}$ and $y_{iva}$, each comprising $M$ channels, are obtained by applying fixed beamforming~\cite{gannot2001signal} and IVA successively to the multi-channel recorded mixtures $y$ , which consist of $P \times M$ channels, which can be illustrated as:
\begin{align}
\label{eq:preprocess}
y_{bf} &= \text{BF}(y), \\
y_{iva} &= \text{IVA}(y_{bf}). 
\end{align}
After applying the short-time Fourier transform (STFT) to obtain complex spectra $Y_{bf}, Y_{iva} \in \mathbb{R}^{T\times F\times M}$, where $T$, $F$ denotes time frames and frequency bins, respectively. The real and imaginary parts are then concatenated along the channel dimension to form input features $R_{bf}, R_{iva} \in \mathbb{R}^{T\times F\times 2M}$. These features are further fed into a dual-encoder, which leverages the guidance of IVA and effectively incorporates spatial prior by modeling features from different data distributions. After obtaining compact latent features, the triple-path modeling module captures structural characteristics between diverse frequency bins, dependencies across different frames, and the correlations between distinct channels, facilitating effective context modeling. Finally, the decoder, mirroring the structure of the encoder, reconstructs the latent features into masks. The predicted masks are element-wise multiplied with the original spectrum to obtain enhanced and separated speech in each speech zone.

\subsection{Unified Streaming and Non-Streaming Architecture}
To create a flexible system capable of handling both streaming and non-streaming scenarios, we design our architecture to seamlessly switch between these modes. To enable streaming functionality, we make specific adjustments to components that depend on future information. In this context, we replace bidirectional recurrent layers and convolution layers with unidirectional recurrent layers and causal convolution layers~\cite{tan2018convolutional}, respectively. In traditional convolutional neural networks, padding is typically applied to both sides of the input to maintain equal lengths of input and output features. This ensures that the convolutional kernel can access historical, current, and future information within its perceptual field. However, in the streaming scenario where future information is not available~\cite{he2019streaming}, causal convolution layers become essential. These layers feature padding only on the left side of the input, effectively excluding future information. To further adapt to streaming requirements, we employ unidirectional recurrent layers. This allows the model to operate without relying on any future information, utilizing only the current frame and the last hidden states as input. These adjustments ensure the system's effectiveness in both streaming and non-streaming modes.

\subsection{Gated Temporal-Frequency Convolutional Block}
\begin{figure}[!t]
    \centering
    \subcaptionbox{Gated Temporal-Frequency Convolutional Block\label{fig:convblock}}{\includegraphics[scale=0.55]{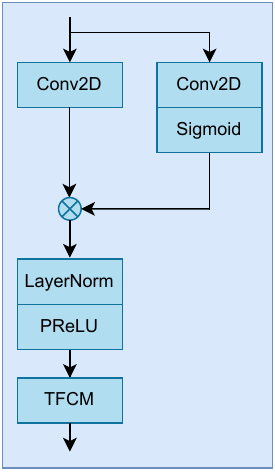}}
    \subcaptionbox{Temporal-Frequency Convolutional Module\label{fig:tfcm}}{\includegraphics[scale=0.55]{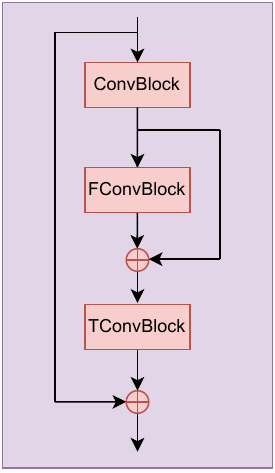}}
    \subcaptionbox{Triple-Path Modeling Module\label{fig:ftlstm}}{\includegraphics[scale=0.55]{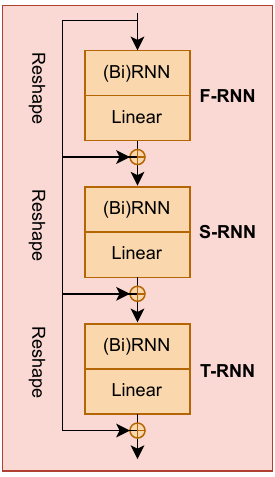}}    
    \caption{Detailed structure of the components in the proposed DualSep. }
    \label{fig:blocks}
\end{figure}

To strengthen feature extraction and context modeling across both temporal and frequency dimensions, we introduce the gated temporal-frequency convolutional block.
As depicted in Fig. \ref{fig:convblock}, this block integrates a gated (transposed) convolution block with several temporal-frequency convolutional modules (TFCM) to form the encoder and decoder. The TFCM, shown in Fig. \ref{fig:tfcm}, consists of three convolutional blocks, each containing a convolution layer, a layer normalization layer, and an activation layer. The key differences among these blocks lie in their kernel size ($K$) and dilation rate ($D$). The first convblock functions as a feature extractor with $K=1$ and $D=1$. The 'FConvBlock' is designed to capture structure patterns across frequency bands by using large $K$ and $D$ along the frequency dimension, whereas the 'TConvBlock' captures context dependency across frames with large $K$ and $D$ along the temporal dimension. Both 'FConvBlock' and 'TConvBlock' use padding to maintain consistent input and output shapes, ensuring causality in the 'TConvBlock' during streaming mode. Skip connections are employed to stabilize the training process~\cite{he2016deep}. The TFCM operates as a dual-path RNN~\cite{luo2020dual}, utilizing fully convolutional layers to achieve effective modeling with low latency.

\subsection{Spatial-Spectral Dual Encoding}
In this section, we explore spatial-spectral dual encoding by introducing a dual encoder framework, combining fixed beamforming and Independent Vector Analysis (IVA) to reduce computational costs and derive spatial prior.

Fixed beamforming is one of the beamforming techniques that only relies on the Directions of Arrival (DOA) or the Relative Transfer Functions (RTFs) of the target source. It can extract the signal of interest from a specific direction, reducing the effects of noise and reverberation in multi-channel signals~\cite{gannot2017consolidated}. This method is suitable when the target direction is known a prior. In our scenario, we assume that the target speech comes directly from the front of the microphone line array in each speech zone. Therefore, we implement the delay-and-sum beamformer~\cite{van1988beamforming} to reduce the number of audio channels, obtaining $y_{bf}$ from $y$. Mathematically, the delay-and-sum beamforming can be expressed as:
\begin{align}
y_{bf_{i}}(t) = \sum_{p=1}^{P} y_{p_{i}}(t - \tau_p), \quad i \in \{1, 2, \ldots, M\}
\end{align}
Here, \( y_{bf_{i}}(t) \) is the beamformed output, \( y_{p_{i}} \) represents the input signal from the \( p \)-th microphone, and \( \tau_p \) is the time delay applied to the \( p \)-th microphone signal to align the signals from the desired direction, and $i$ denotes the index of each speech zone.

Independent Vector Analysis (IVA) is an extension of Independent Component Analysis (ICA)~\cite{lee1998independent} for separating mixed sources in multi-channel signals. 
In our system, we leverage IVA as preprocessing to derive spatial prior. Given the multi-channel complex spectrograms $Y_{bf} \in \mathbb{R}^{T\times F\times M}$, the unmixing matrices $\mathbf{W}$ are initialized to the identity matrix $\mathbf{I}$ for each frequency bin and then used to decorrelate $Y_{bf}$. The unmixing matrices are updated iteratively using a gradient-based method to obtain the separated spectrograms $Y_{iva}$. $\mathbf{g(\cdot)}$ is a non-linear function to ensure statistical independence among the sources. The learning rate is denoted by $\eta$. This process can be shown as:
\begin{align}
\label{eq:iva}
\mathbf{W}_i &= \mathbf{I}, \quad i \in \{1, 2, \ldots, F\}, \\
Y_{iva_t} &= \mathbf{W} Y_{bf_t}, \quad t \in \{1, 2, \ldots, T\}, \\
\mathbf{W}_{i+1} &= \mathbf{W}_i - \eta \left( \frac{1}{T} \sum_{t=1}^{T} \mathbf{g}(Y_{iva_t}) Y_{iva_t}^H \mathbf{W}_i^{-H} - \mathbf{I} \right). 
\end{align}

After obtaining  $y_{bf}$ and $y_{iva}$, we employ a dual encoder framework~\cite{chidambaram2018learning} to separately model the spatial and spectral components present in the multichannel scenarios.
The spatial encoder is designed to capture spatial cues, while the spectral encoder focuses on preserving spectral information. The spectral encoder processes multichannel signals $R_{bf}$ as input, while the spatial encoder works with multichannel signals obtained after applying IVA, denoted as $R_{iva}$. This dual encoder strategy allows the comprehensive modeling of both spatial and spectral aspects of multichannel signals, capturing their unique characteristics separately.

\subsection{Triple-Path Modeling Module}
The triple-path modeling module is composed of three RNN blocks, as illustrated in Fig. \ref{fig:ftlstm}, integrating frequency-focused, spatial-focused, and temporal-focused RNN blocks with skip connections to comprehensive modeling. Each block combines a (bidirectional) recurrent neural network~\cite{memory2010long} with a linear layer to capture distinct types of information. Skip connections are incorporated to alleviate potential information loss~\cite{he2016deep}. Given input features $ R \in \mathbb{R}^{\text{T} \times \text{F} \times \text{D}}$, where $\text{T}$, $\text{F}$, and $\text{D}$ represent time frames, frequency bins, and feature dimensions, respectively, the first block, denoted as ``$\text{F-RNN}$'', focuses on exploiting single-channel spectral information along the frequency axis. Employing a bidirectional RNN allows this block to emphasize frequency dependencies in each frame, effectively capturing spectral patterns crucial for distinguishing between speech and noise. The second block, labeled ``$\text{S-RNN}$'', targets the exploitation of multichannel spatial information. This block endeavors to learn the correlation of spatial cues, acting as a spatial clustering mechanism due to the distinct differences in signals between each speech zone. It implicitly separates different speakers in different speech zones. The third block, named ``$\text{T-RNN}$'', capitalizes on the full-band pattern and its temporal contexts. It collects and integrates information obtained from the previous blocks, offering a comprehensive understanding of both spectral and spatial characteristics. Linear layers are used to align feature dimensions while reducing the number of parameters and computational complexity. Mathematically, this can be expressed as:
\begin{align*}
R_{\text{F-RNN}} &= \text{F-RNN}_{\text{bi}}(R) + R \\
R_{\text{S-RNN}} &= \text{S-RNN}(R_{\text{F-RNN}}) + R \\
R_{\text{T-RNN}} &= \text{T-RNN}(R_{\text{S-RNN}}) + R
\end{align*}

\section{Experiment}
\label{sec:exp}

\subsection{Datasets and Evaluation Metrics}
\textbf{Datasets} We simulate multi-speaker in-car datasets using AISHELL-1 (clean speech corpus)~\cite{bu2017aishell} and DNS challenge 5 (noise corpus)~\cite{dubey2023icassp}. Each speech zone is equipped with a 4-microphone line array, with microphones spaced 2cm apart. We generate a total of 10K multichannel room impulse responses(RIRs)  using image-source method\footnote{\url{https://github.com/DavidDiazGuerra/gpuRIR/}}
, covering different sizes of standard vehicle cabins. The width, length, and height of the cabins fall within the range of [1.5m, 1.7m], [2.3m, 2.5m], and [1.2m, 1.5m], respectively. One to six passengers are speaking simultaneously, with the flexibility to move within their respective speech zones. Each multi-channel RIR is a set consisting of RIRs from six speakers (one in each zone) and background noise. The reverberation time (RT60) ranges between [0.3s, 0.7s] across configurations. We randomly select RIRs to simulate the in-car dataset. In addition, we include diffused noise with signal-to-distortion ratio (SDR) ranging from [-20, 15] dB. A total of 100K, 5K, and 1K utterances are generated for the ‘Train’, ‘Dev’, and ‘Test’ datasets. Importantly, there are no overlaps in noise types and RIRs across the ‘Train’, ‘Dev’, and ‘Test’ datasets to ensure unbiased evaluation.

\textbf{Evaluation Metrics} The objective metrics we adopt include PESQ, Si-SNR, and Character Error Rate (CER) to comprehensively evaluate different systems from multiple aspects such as speech quality, perceptual performance, and intelligibility. We also compare the computational complexity and cost required for each system by computing Real-Time Factor (RTF).

\subsection{Experimental Setup}
\textbf{Data Preprocessing} In our scenario, each speech zone is equipped with a 4-microphone line array, resulting in initial input signals $y$ comprising 24 channels in total. Fixed beamforming is utilized in each speech zone to enhance signal quality and reduce the computational cost, yielding 6-channel signals $y_{bf}$. Subsequently, we apply IVA to $y_{bf}$ as preprocessing to roughly separate different speakers, resulting in $y_{iva}$, which serves as spatial prior.

\textbf{Feature Extraction and Network Configuration} A 512-point FFT is applied along with 32 ms Hann window and 16 ms step size to extract the audio features, resulting in $Y_{bf}$, $Y_{iva}$. For the network configuration, the spatial encoder and spectral encoder each consist of 5 gated temporal-frequency convolutional blocks. Each block comprises 4 layers of TFCM, with channels set to \{12, 12, 24, 48, 64\} in each block. The bottleneck comprises 2 layers of triple-path modeling module, with the hidden dimension set to 64. The spectral decoder mirrors the settings of the spectral encoder. Kernel size and stride are set to (3,2) and (2,1) in the frequency and temporal dimensions, respectively. In streaming mode, necessary paddings and non-bidirectional RNNs are employed.

\textbf{Baselines and Training Procedure} In our study, we compare our proposed system to several competitive baselines, including (i) MIMO UNet~\cite{ronneberger2015u}, (ii) Multi-ch ConvTasNet~\cite{luo2019conv}, (iii) ADL-MVDR~\cite{zhang2021adl}, (iv) McNet \footnote{\url{https://github.com/Audio-WestlakeU/McNet/}}~\cite{yang2023mcnet} and (v) ZoneFormer~\cite{xu2023zoneformer}. For the comparison system, the input is $R \in \mathbb{R}^{T\times F\times 4M}$ obtained by concatenating $R_{bf}$,$R_{iva}$. 
All systems are trained on 4-second chunks using the AdamW optimizer~\cite{loshchilov2017decoupled}
, with a batch size varied between 8 and 16 to optimize the time-domain Scale-Invariant Source-to-Noise Ratio (SiSNR) and minimize the frequency-domain Mean Square Error (MSE) between estimated speaker speech and corresponding reverberant clean speech. The initial learning rate is set to 1e-3, with gradient norm clipped at a maximum norm of 5.
Additionally, we also conduct experiments with the DSP methods used in the proposed system: fixed beamforming (BF), and fixed beamforming combined with Independent Vector Analysis (BF+IVA).

\section{Results and Discussion}
\label{sec:result}

\subsection{Results}
Table \ref{table:baseline} presents the experimental results for different speech separation systems on the test set, covering various evaluation metrics. These metrics include parameters and Real-Time Factor (RTF) to evaluate model complexity and computational efficiency, with causality measured for real-time applicability. Additionally, SiSNR and PESQ assess speech quality and subjective perception, while CER \footnote{\url{https://github.com/wenet-e2e/wenet/tree/main/examples/aishell/s0/}} reflects speech intelligibility, crucial for downstream tasks.
The unit for parameters is in millions (M), CER is expressed as a percentage (\%), and SiSNR is measured in dB.

\begin{table}[htbp]
    \caption{Experimental results for different speech separation systems across objective evaluation metrics}
    \centering
    \captionsetup{font=footnotesize} 
    \setlength{\tabcolsep}{1.5pt} 
    \renewcommand{\arraystretch}{0.9} 
    \footnotesize 
    \begin{tabular}{
            l
            S[table-format=2.2]
            c
            c 
            S[table-format=1.2]
            c 
            S[table-format=1.2]            
        }
        \toprule
        System             & {\#Para(M)} & Casual & \multicolumn{1}{c}{SiSNR $\uparrow$} & {PESQ$\uparrow$} & \multicolumn{1}{c}{CER $\downarrow$} & {RTF$\downarrow$}  \\
        \midrule
        Unprocessed         & {-}       & {-}    & -1.83        & 1.31   & 89.32      & {-}    \\
        \midrule
        BF         & {-}       & {-}    & 1.61        & 1.61   & 86.13       & {-}    \\
        BF+IVA         & {-}       & {-}    & 3.97        & 1.73   & 79.96    & {-}    \\
        \midrule        
        MIMO-UNet           & 1.47     & \checkmark & 7.93   & 2.45   & 40.98      & 0.30   \\
        Multi-ch ConvTasNet & 15.10    & \texttimes & 8.73   & 2.88   & 31.37      & 1.34   \\
        ADL-MVDR            & 6.63     & \texttimes & 8.98   & 2.99   & 15.01      & 0.73   \\
        McNet               & 1.91     & \checkmark & 9.33   & 3.07   & 11.33      & 0.60   \\
        ZoneFormer               & 1.85     & \checkmark & 10.31   & 2.95   & 9.33      & 0.75   \\        
        \midrule
        \textbf{Prop.} DualSep-S         & 1.05    & \texttimes & 13.50  & 3.87   &  7.65       & 0.45   \\
        \textbf{Prop.} DualSep-S     & 0.83     & \checkmark & 12.08  & 3.35   & 8.44       & \textbf{0.39}   \\
        \textbf{Prop.} DualSep-L         & 1.12     & \texttimes & \textbf{13.97}  & \textbf{3.90}   &  \textbf{6.99}       & 0.48   \\        
        \textbf{Prop.} DualSep-L     & 0.91    & \checkmark & 12.92  & 3.35   & 7.90       & 0.41   \\
        \midrule
        Reverb. Clean Ref.  & {-}       & {-}    & {$\infty$}     & 4.40    & 4.36       & {-}    \\
        \bottomrule
    \end{tabular}
    \label{table:baseline}
\end{table}

\textbf{The performance of baseline systems} 
The experimental results show the effectiveness of the strong baseline systems. Across a complex and challenging multi-talker test set,  these baselines exhibit significant improvements, with SiSNR increasing by at least 8dB and CER improving by a minimum of 40\%, improving both objective perception and intelligibility.

\textbf{Comparison with baseline systems}
The proposed system outperforms the baseline systems across objective metrics, including SiSNR improvement, PESQ score, and CER reduction. Notably, the efficiency of the dual-encoder system stands out, achieving superior performance with minimal parameters and the smallest RTF. By inputting different signals $R_{bf}$ and $R_{iva}$ into the \textit{spectral encoder} and \textit{spatial encoder} respectively, the dual-encoder architecture effectively integrates spatial and spectral information, resulting in enhanced speech separation performance compared to baseline systems.

\textbf{Comparison with different variants of the proposed system} 
DualSep-S and DualSep-L differ in their fusion methods for spatial and spectral cues. In DualSep-S, the latent representations from the \textit{spatial encoder} and \textit{spectral encoder} are added up, while in DualSep-L, they are concatenated. Although DualSep-L performs slightly better, the concatenation method better preserve spatial and spectral information. However, this comes at the cost of increased parameters and RTF due to larger hidden dimensions. Additionally, performance comparisons are made between streaming and non-streaming modes, with the proposed system demonstrating comparable performance in streaming mode with smaller parameters and RTF, which is promising for real-time applications.

\textbf{Analysis of the effects of the DSP methods used in the proposed system} 
Experimental results reveal that using fixed beamforming (BF) alone leads to very limited improvement. This is expected as we employed a simple delay-and-sum beamforming method primarily to reduce the number of audio channels and computational load. Although the improvements in metrics is modest, BF retains rich spectral information. When combining BF with Independent Vector Analysis (BF+IVA), there is a noticeable improvement, with a nearly 10\% reduction in CER. This aligns with our expectations since IVA, a classical blind source separation algorithm, can separate speech from different speakers to their respective zones, providing useful spatial prior. After pre-processing with DSP methods, the outputs $y_{bf}$ and  $y_{iva}$ are fed into the proposed DualSep system. The results show significant improvements across all metrics, with marked enhancements in both listening quality and recognition rates compared to baseline systems, demonstrating the effectiveness of the proposed method. 

\subsection{Ablation Study}

We conduct an ablation study to verify the effectiveness of the dual-encoder architecture. With the rest remaining unchanged, \textit{Spectral Encoder Only} uses $R_{bf}$ as input, \textit{Spatial Encoder Only} uses $R_{iva}$ as input, and \textit{Combined Encoder} uses $R$ (obtained by concatenating $R_{bf}$, $R_{iva}$ along the channel dimension.) as input with single encoder.

\begin{table}[]
    \caption{Comparison of different encoder configurations}
    \centering
    \captionsetup{font=footnotesize} 
    \setlength{\tabcolsep}{1.5pt} 
    \renewcommand{\arraystretch}{1.5} 
    \footnotesize 
\begin{tabular}{ccccccc}
\hline
System               & {\#Param} & Casual & {SiSNR $\uparrow$} & {PESQ$\uparrow$} & {CER $\downarrow$} & {RTF$\downarrow$}  \\ \hline
\textit{Dual Encoder} & 0.83   & \checkmark     & 12.08       & 3.35 & 8.44    & 0.39 \\ \hline
\textit{Spectral Encoder Only} & 0.65   & \checkmark     & 9.93       & 2.97 & 13.66    & 0.33 \\
\textit{Spatial Encoder Only}  & 0.65   & \checkmark      & 7.21       & 2.19 & 31.33    & 0.33 \\
\textit{Combined Encoder}      & 0.71   & \checkmark      & 10.01      & 3.00 & 11.50    & 0.37 \\ \hline
\end{tabular}
\label{table:ablation}
\end{table}

\textbf{Comparison with single encoder} Table \ref{table:ablation} highlights a notable decline in performance metrics for single encoder configurations compared to the dual-encoder architecture, with only a marginal reduction in model complexity and computational overhead. For the \textit{Spectral Encoder Only} configuration, the decline in performance can be attributed to its reliance solely on $R_{bf}$ without leveraging spatial priors, it has to implicitly learn to separate different speakers into distinct speech zones, leading to performance degradation. Similarly, the \textit{Spatial Encoder Only} configuration suffers from compromised signal quality and potential loss of crucial information introduced by IVA. Despite utilizing spatial priors through $R_{iva}$, the absence of complete spectral information may result in degraded signal quality and reduced recognition rates.

\begin{figure}[!t]
    \centering
    \begin{tabular}{ccc}
        \subcaptionbox{Speech Zone 1  \label{fig:spk1}}{\includegraphics[width=0.3\linewidth]{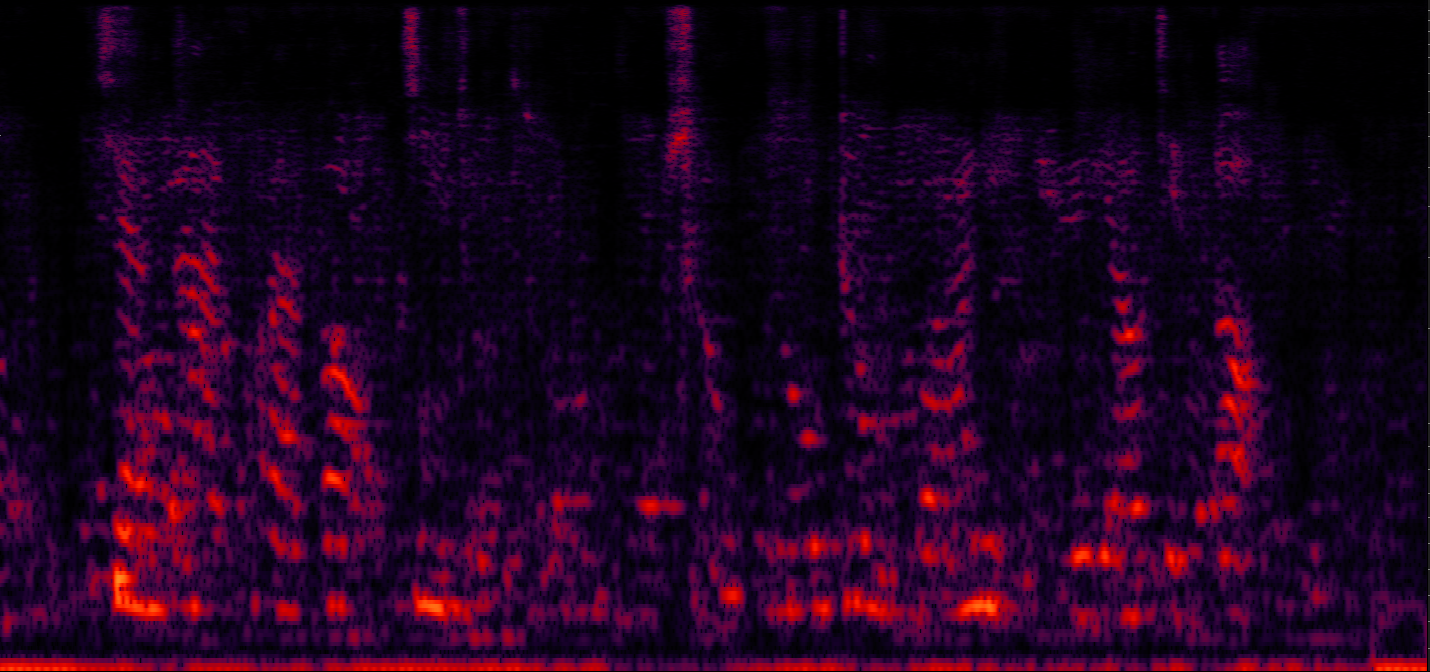}} &
        \subcaptionbox{Speech Zone 2  \label{fig:spk2}}{\includegraphics[width=0.3\linewidth]{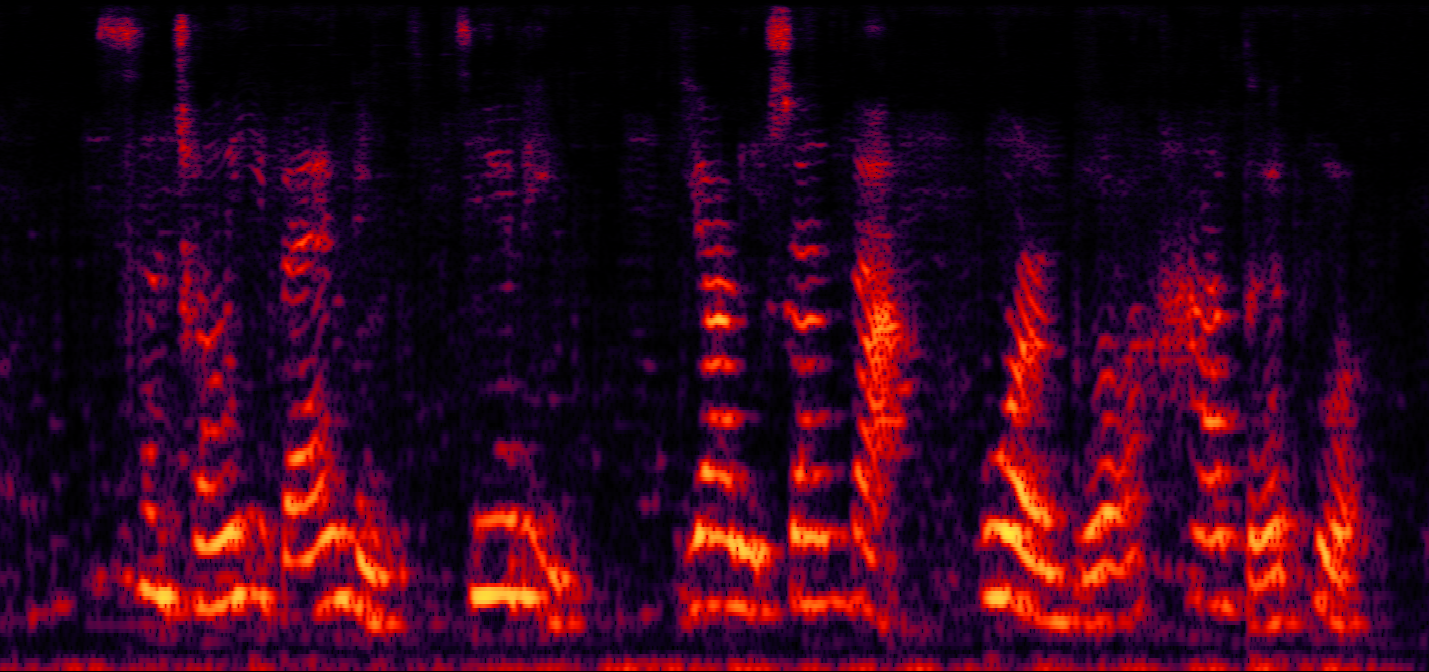}} &
        \subcaptionbox{Speech Zone 3  \label{fig:spk3}}{\includegraphics[width=0.3\linewidth]{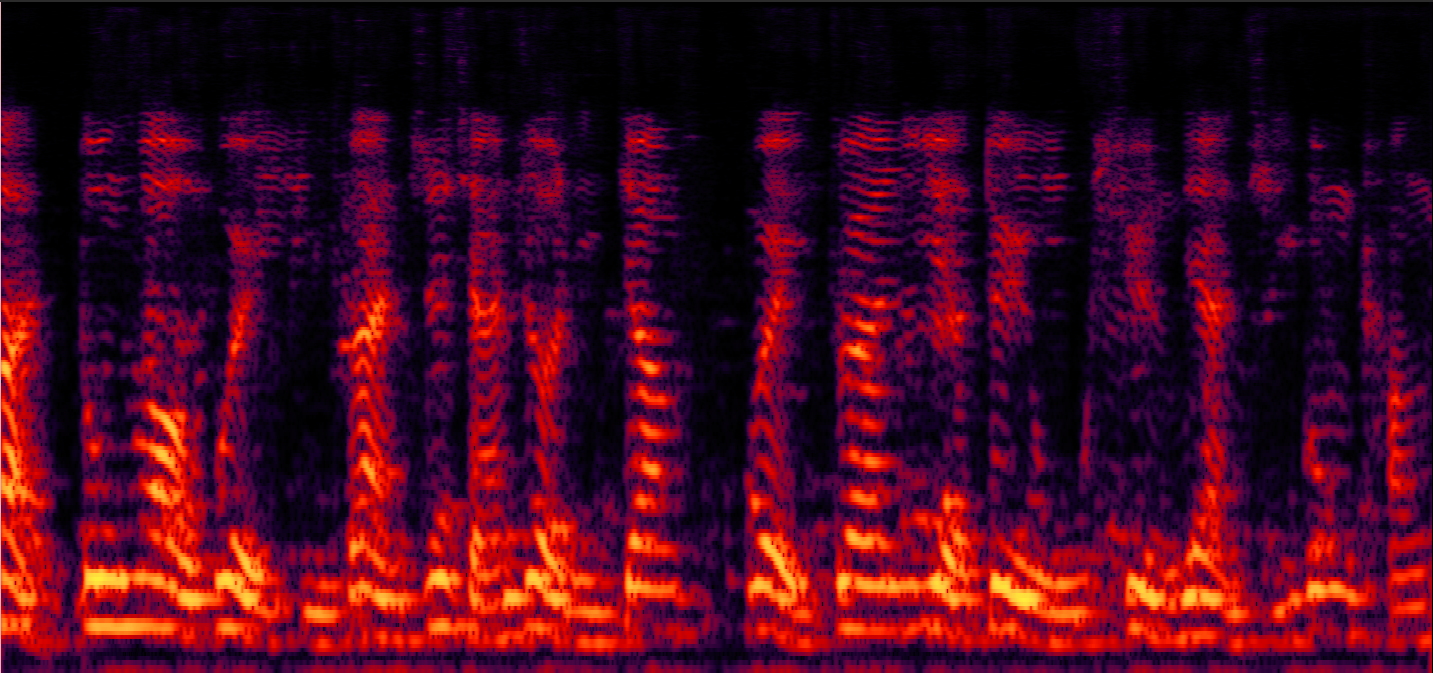}} \\
        \subcaptionbox{Speech Zone 4  \label{fig:spk4}}{\includegraphics[width=0.3\linewidth]{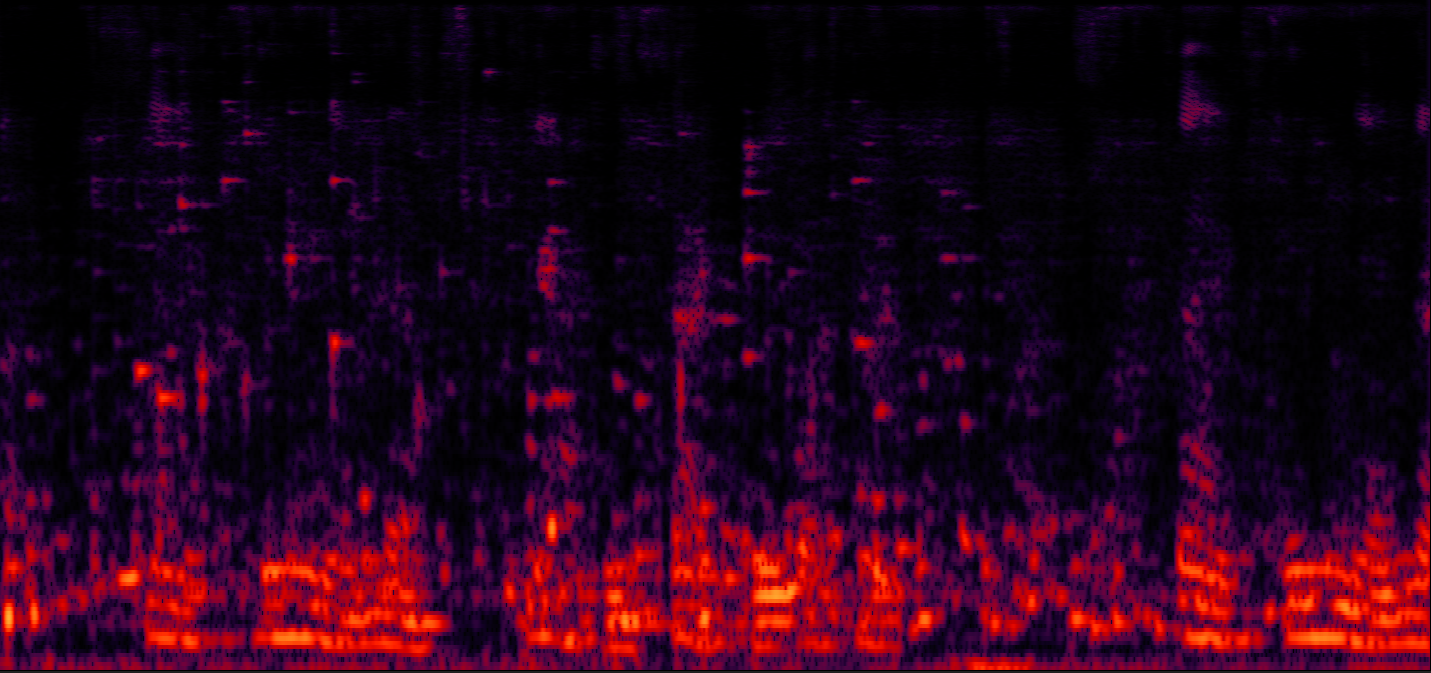}} &
        \subcaptionbox{Speech Zone 5  \label{fig:spk5}}{\includegraphics[width=0.3\linewidth]{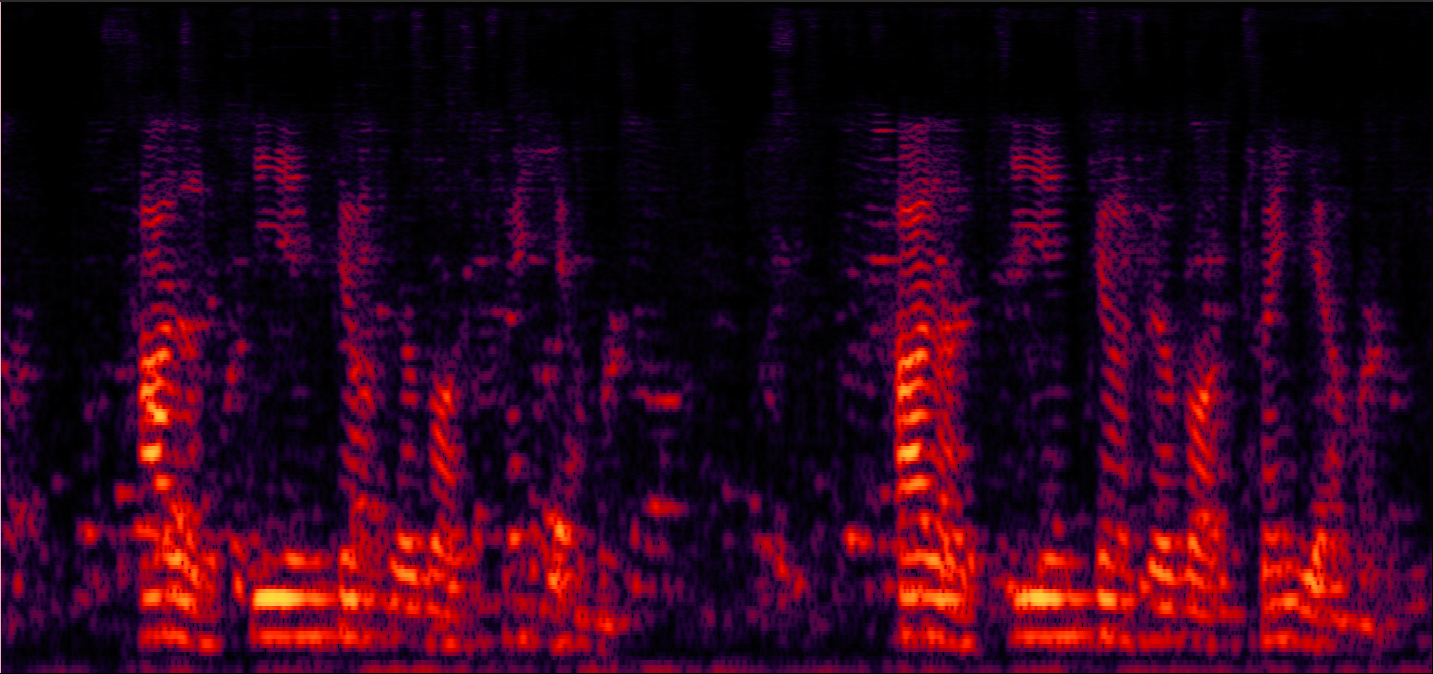}} &
        \subcaptionbox{Speech Zone 6  \label{fig:spk6}}{\includegraphics[width=0.3\linewidth]{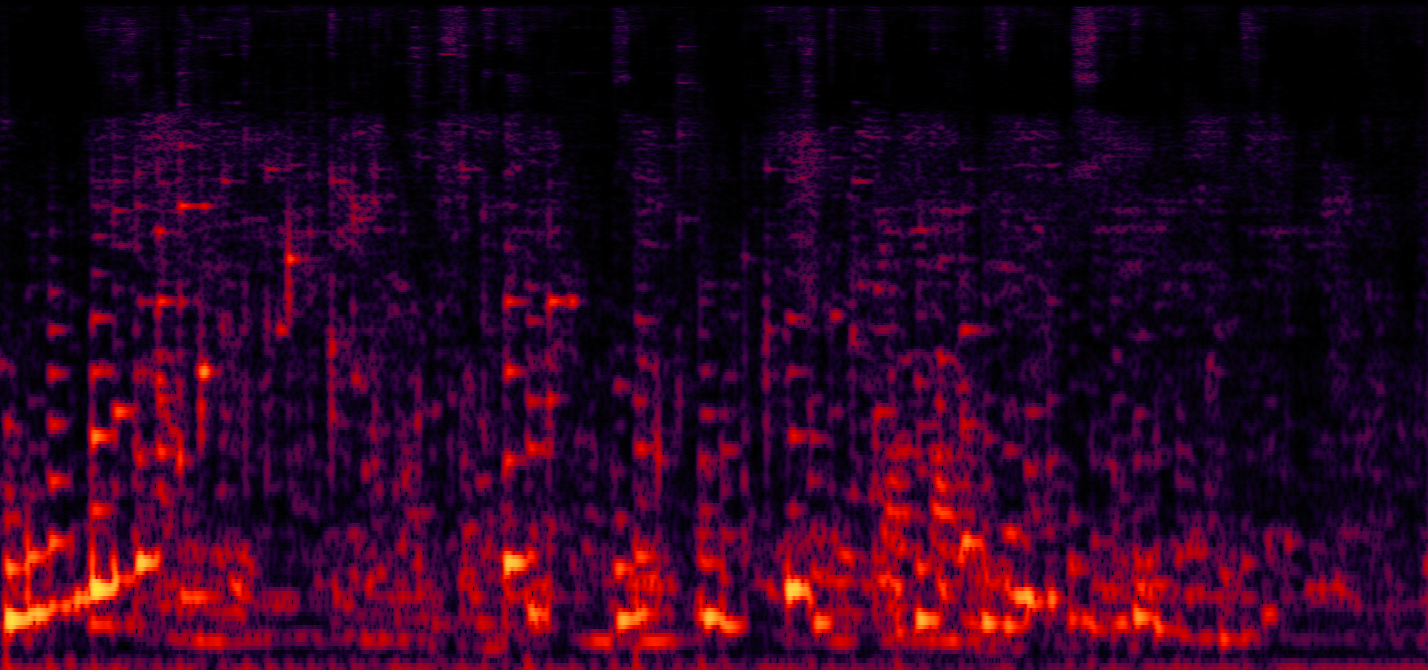}} \\
        \multicolumn{3}{c}{
            \subcaptionbox{Mixture \label{fig:mix}}{\includegraphics[width=0.4\linewidth]{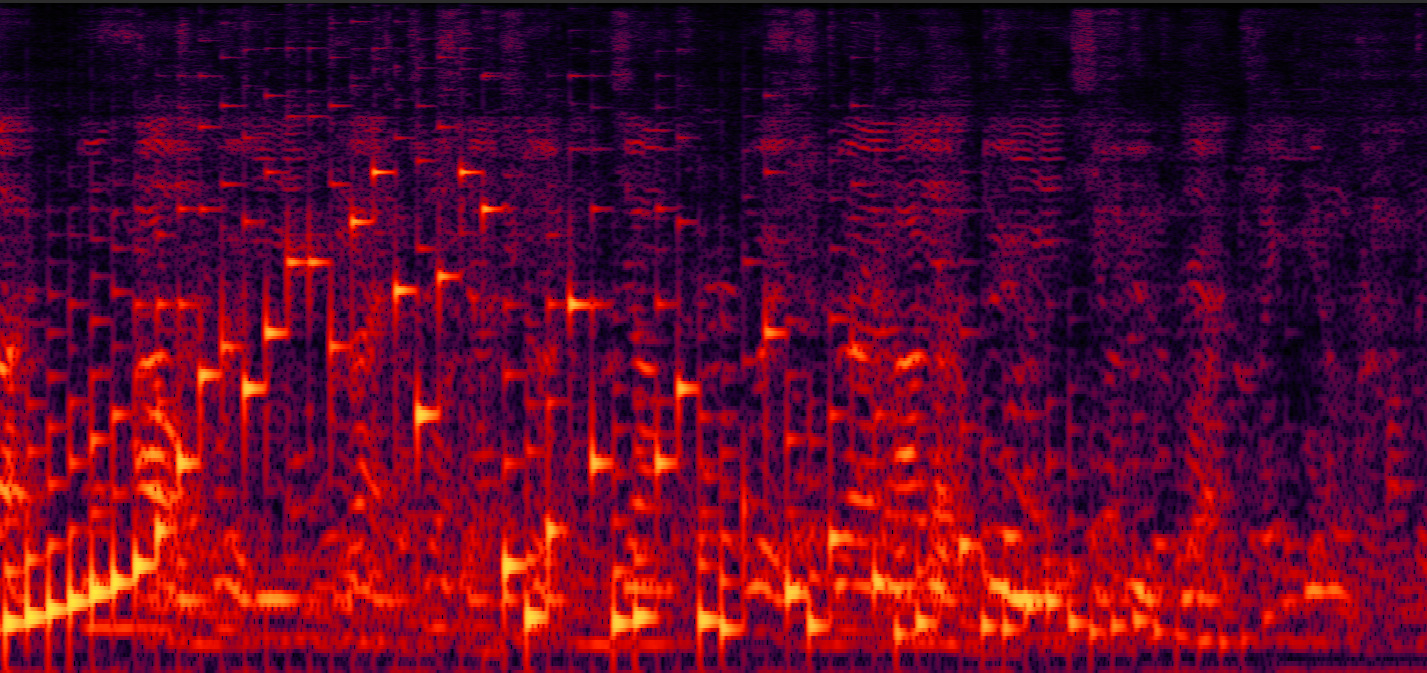}}
        }
    \end{tabular}
    \caption{Separated spectrograms of six speech zones by the proposed system. We only select the spectrogram of one channel from the multi-channel mixture for demonstration.}
    \label{fig:demo}
\end{figure}

\textbf{Comparison with combined encoder}
The performance gap between \textit{Dual Encoder} and \textit{Combined Encoder} underscores the effectiveness of the dual-encoder architecture. \textit{Combined Encoder} merges inputs from $R_{bf}$ and $R_{iva}$ but does not decouple the spatial and spectral information effectively, resulting in performance that is better than single encoders but still inferior to \textit{Dual Encoder}. The dual-encoder architecture's advantage lies in its ability to separately model spatial and spectral components, enabling comprehensive representation and mitigating interference between different types of information. Fig. \ref{fig:demo} is a visualization of the spectrograms of speech separated by the proposed system from a mixture when all speech zones speak simultaneously.



\section{Conclusion}
In this paper, we propose a lightweight framework cascading digital signal processing and neural networks, utilizing dual encoders to effectively leverage spatial prior for intelligent human-vehicle interaction. Our system offers a unified solution for both streaming and non-streaming scenarios, addressing the challenge of separating speech in complex, multi-talker in-car environments. With only 0.83M model parameters and a real-time factor of 0.39 on Intel Core i7 (2.6GHz) CPU, our system demonstrates its superiority across various evaluation metrics, showcasing its potential for real-time applications.



\bibliographystyle{IEEEbib}
\bibliography{mybib}

\begin{thebibliography}{10}

\bibitem{zhang2022end}
Wangyou Zhang, Xuankai Chang, Christoph Boeddeker, Tomohiro Nakatani, Shinji Watanabe, and Yanmin Qian,
\newblock ``End-to-end dereverberation, beamforming, and speech recognition in a cocktail party,''
\newblock {\em IEEE/ACM Transactions on Audio, Speech, and Language Processing}, vol. 30, pp. 3173--3188, 2022.

\bibitem{wang2020complex}
Zhong-Qiu Wang, Peidong Wang, and DeLiang Wang,
\newblock ``Complex spectral mapping for single-and multi-channel speech enhancement and robust asr,''
\newblock {\em IEEE/ACM transactions on audio, speech, and language processing}, vol. 28, pp. 1778--1787, 2020.

\bibitem{kanda2020joint}
Naoyuki Kanda, Yashesh Gaur, Xiaofei Wang, Zhong Meng, Zhuo Chen, Tianyan Zhou, and Takuya Yoshioka,
\newblock ``Joint speaker counting, speech recognition, and speaker identification for overlapped speech of any number of speakers,''
\newblock in {\em Interspeech}, 10 2020, pp. 36--40.

\bibitem{pickering2007research}
Carl~A Pickering, Keith~J Burnham, and Michael~J Richardson,
\newblock ``A research study of hand gesture recognition technologies and applications for human vehicle interaction,''
\newblock in {\em 2007 3rd Institution of Engineering and Technology conference on automotive electronics}. IET, 2007, pp. 1--15.

\bibitem{rahmati2020game}
Yalda Rahmati, Alireza Talebpour, Archak Mittal, and James Fishelson,
\newblock ``Game theory-based framework for modeling human--vehicle interactions on the road,''
\newblock {\em Transportation research record}, vol. 2674, no. 9, pp. 701--713, 2020.

\bibitem{capallera2022human}
Marine Capallera, Leonardo Angelini, Quentin Meteier, Omar Abou~Khaled, and Elena Mugellini,
\newblock ``Human-vehicle interaction to support driver's situation awareness in automated vehicles: A systematic review,''
\newblock {\em IEEE Transactions on Intelligent Vehicles}, 2022.

\bibitem{murali2022intelligent}
Prajval~Kumar Murali, Mohsen Kaboli, and Ravinder Dahiya,
\newblock ``Intelligent in-vehicle interaction technologies,''
\newblock {\em Advanced Intelligent Systems}, vol. 4, no. 2, pp. 2100122, 2022.

\bibitem{xu2021generalized}
Yong Xu, Zhuohuang Zhang, Meng Yu, Shi-Xiong Zhang, and Dong Yu,
\newblock ``Generalized spatio-temporal rnn beamformer for target speech separation,''
\newblock in {\em Interspeech}, 08 2021, pp. 3076--3080.

\bibitem{wang2021multi}
Zhong-Qiu Wang, Peidong Wang, and DeLiang Wang,
\newblock ``Multi-microphone complex spectral mapping for utterance-wise and continuous speech separation,''
\newblock {\em IEEE/ACM transactions on audio, speech, and language processing}, vol. 29, pp. 2001--2014, 2021.

\bibitem{zhang2022all}
Zhuohuang Zhang, Takuya Yoshioka, Naoyuki Kanda, Zhuo Chen, Xiaofei Wang, Dongmei Wang, and Sefik~Emre Eskimez,
\newblock ``All-neural beamformer for continuous speech separation,''
\newblock in {\em ICASSP 2022-2022 IEEE International Conference on Acoustics, Speech and Signal Processing (ICASSP)}. IEEE, 2022, pp. 6032--6036.

\bibitem{zhang2021adl}
Zhuohuang Zhang, Yong Xu, Meng Yu, Shi-Xiong Zhang, Lianwu Chen, and Dong Yu,
\newblock ``Adl-mvdr: All deep learning mvdr beamformer for target speech separation,''
\newblock in {\em ICASSP 2021-2021 IEEE International Conference on Acoustics, Speech and Signal Processing (ICASSP)}. IEEE, 2021, pp. 6089--6093.

\bibitem{van2009speech}
Tim Van~den Bogaert, Simon Doclo, Jan Wouters, and Marc Moonen,
\newblock ``Speech enhancement with multichannel wiener filter techniques in multimicrophone binaural hearing aids,''
\newblock {\em The Journal of the Acoustical Society of America}, vol. 125, no. 1, pp. 360--371, 2009.

\bibitem{kothapally2023deep}
Vinay Kothapally, Yong Xu, Meng Yu, Shi-Xiong Zhang, and Dong Yu,
\newblock ``Deep neural mel-subband beamformer for in-car speech separation,''
\newblock in {\em ICASSP 2023-2023 IEEE International Conference on Acoustics, Speech and Signal Processing (ICASSP)}. IEEE, 2023, pp. 1--5.

\bibitem{zhao2012fast}
Shengkui Zhao and Douglas~L Jones,
\newblock ``A fast-converging adaptive frequency-domain mvdr beamformer for speech enhancement.,''
\newblock in {\em Interspeech}, 2012, pp. 1930--1933.

\bibitem{xiao2017time}
Xiong Xiao, Shengkui Zhao, Douglas~L Jones, Eng~Siong Chng, and Haizhou Li,
\newblock ``On time-frequency mask estimation for mvdr beamforming with application in robust speech recognition,''
\newblock in {\em 2017 IEEE International Conference on Acoustics, Speech and Signal Processing (ICASSP)}. IEEE, 2017, pp. 3246--3250.

\bibitem{boeddeker2018exploring}
Christoph Boeddeker, Hakan Erdogan, Takuya Yoshioka, and Reinhold Haeb-Umbach,
\newblock ``Exploring practical aspects of neural mask-based beamforming for far-field speech recognition,''
\newblock in {\em 2018 IEEE international conference on acoustics, speech and signal processing (ICASSP)}. IEEE, 2018, pp. 6697--6701.

\bibitem{xu2020neural}
Yong Xu, Meng Yu, Shi-Xiong Zhang, Lianwu Chen, Chao Weng, Jianming Liu, and Dong Yu,
\newblock ``Neural spatio-temporal beamformer for target speech separation,''
\newblock {\em arXiv preprint arXiv:2005.03889}, 2020.

\bibitem{gannot2017consolidated}
Sharon Gannot, Emmanuel Vincent, Shmulik Markovich-Golan, and Alexey Ozerov,
\newblock ``A consolidated perspective on multimicrophone speech enhancement and source separation,''
\newblock {\em IEEE/ACM Transactions on Audio, Speech, and Language Processing}, vol. 25, no. 4, pp. 692--730, 2017.

\bibitem{kim2006independent}
Taesu Kim, Intae Lee, and Te-Won Lee,
\newblock ``Independent vector analysis: Definition and algorithms,''
\newblock in {\em 2006 Fortieth Asilomar Conference on Signals, Systems and Computers}. IEEE, 2006, pp. 1393--1396.

\bibitem{chidambaram2018learning}
Muthu Chidambaram, Yinfei Yang, Daniel Cer, Steve Yuan, Yunhsuan Sung, Brian Strope, and Ray Kurzweil,
\newblock ``Learning cross-lingual sentence representations via a multi-task dual-encoder model,''
\newblock in {\em Proceedings of the 4th Workshop on Representation Learning for NLP (RepL4NLP-2019)}, 2019, pp. 250--259.

\bibitem{gannot2001signal}
Sharon Gannot, David Burshtein, and Ehud Weinstein,
\newblock ``Signal enhancement using beamforming and nonstationarity with applications to speech,''
\newblock {\em IEEE Transactions on Signal Processing}, vol. 49, no. 8, pp. 1614--1626, 2001.

\bibitem{tan2018convolutional}
Ke~Tan and DeLiang Wang,
\newblock ``A convolutional recurrent neural network for real-time speech enhancement.,''
\newblock in {\em Interspeech}, 2018, vol. 2018, pp. 3229--3233.

\bibitem{he2019streaming}
Yanzhang He, Tara~N Sainath, Rohit Prabhavalkar, Ian McGraw, Raziel Alvarez, Ding Zhao, David Rybach, Anjuli Kannan, Yonghui Wu, Ruoming Pang, et~al.,
\newblock ``Streaming end-to-end speech recognition for mobile devices,''
\newblock in {\em ICASSP 2019-2019 IEEE International Conference on Acoustics, Speech and Signal Processing (ICASSP)}. IEEE, 2019, pp. 6381--6385.

\bibitem{he2016deep}
Kaiming He, Xiangyu Zhang, Shaoqing Ren, and Jian Sun,
\newblock ``Deep residual learning for image recognition,''
\newblock in {\em Proceedings of the IEEE conference on computer vision and pattern recognition}, 2016, pp. 770--778.

\bibitem{luo2020dual}
Yi~Luo, Zhuo Chen, and Takuya Yoshioka,
\newblock ``Dual-path rnn: efficient long sequence modeling for time-domain single-channel speech separation,''
\newblock in {\em ICASSP 2020-2020 IEEE International Conference on Acoustics, Speech and Signal Processing (ICASSP)}. IEEE, 2020, pp. 46--50.

\bibitem{van1988beamforming}
Barry~D Van~Veen and Kevin~M Buckley,
\newblock ``Beamforming: A versatile approach to spatial filtering,''
\newblock {\em IEEE assp magazine}, vol. 5, no. 2, pp. 4--24, 1988.

\bibitem{lee1998independent}
Te-Won Lee and Te-Won Lee,
\newblock {\em Independent component analysis},
\newblock Springer, 1998.

\bibitem{memory2010long}
Long Short-Term Memory,
\newblock ``Long short-term memory,''
\newblock {\em Neural computation}, vol. 9, no. 8, pp. 1735--1780, 2010.

\bibitem{bu2017aishell}
Hui Bu, Jiayu Du, Xingyu Na, Bengu Wu, and Hao Zheng,
\newblock ``Aishell-1: An open-source mandarin speech corpus and a speech recognition baseline,''
\newblock in {\em 2017 20th conference of the oriental chapter of the international coordinating committee on speech databases and speech I/O systems and assessment (O-COCOSDA)}. IEEE, 2017, pp. 1--5.

\bibitem{dubey2023icassp}
Harishchandra Dubey, Ashkan Aazami, Vishak Gopal, Babak Naderi, Sebastian Braun, Ross Cutler, Alex Ju, Mehdi Zohourian, Min Tang, Hannes Gamper, et~al.,
\newblock ``Icassp 2023 deep speech enhancement challenge,''
\newblock {\em arXiv preprint arXiv:2303.11510}, 2023.

\bibitem{ronneberger2015u}
Olaf Ronneberger, Philipp Fischer, and Thomas Brox,
\newblock ``U-net: Convolutional networks for biomedical image segmentation,''
\newblock in {\em Medical Image Computing and Computer-Assisted Intervention--MICCAI 2015: 18th International Conference, Munich, Germany, October 5-9, 2015, Proceedings, Part III 18}. Springer, 2015, pp. 234--241.

\bibitem{luo2019conv}
Yi~Luo and Nima Mesgarani,
\newblock ``Conv-tasnet: Surpassing ideal time--frequency magnitude masking for speech separation,''
\newblock {\em IEEE/ACM transactions on audio, speech, and language processing}, vol. 27, no. 8, pp. 1256--1266, 2019.

\bibitem{yang2023mcnet}
Yujie Yang, Changsheng Quan, and Xiaofei Li,
\newblock ``Mcnet: Fuse multiple cues for multichannel speech enhancement,''
\newblock in {\em ICASSP 2023-2023 IEEE International Conference on Acoustics, Speech and Signal Processing (ICASSP)}. IEEE, 2023, pp. 1--5.

\bibitem{xu2023zoneformer}
Yong Xu, Vinay Kothapally, Meng Yu, Shi-Xiong Zhang, and Dong Yu,
\newblock ``Zoneformer: On-device neural beamformer for in-car multi-zone speech separation, enhancement and echo cancellation,''
\newblock Interspeech, 2023.

\bibitem{loshchilov2017decoupled}
Ilya Loshchilov and Frank Hutter,
\newblock ``Decoupled weight decay regularization,''
\newblock in {\em International Conference on Learning Representations}, 2018.

\end{thebibliography}

\end{document}